# Distributed Power Control for Delay Optimization in Energy Harvesting Cooperative Relay Networks

V. Hakami and M. Dehghan, *Member, IEEE*

*Abstract*— We consider cooperative communications with energy harvesting (EH) relays, and develop a distributed power control mechanism for the relaying terminals. Unlike prior art which mainly deal with single-relay systems with saturated traffic flow, we address the case of bursty data arrival at the source cooperatively forwarded by multiple half-duplex EH relays. We aim at optimizing the long-run average delay of the source packets under the energy neutrality constraint on power consumption of each relay. While EH relay systems have been predominantly optimized using either offline or online methodologies, we take on a more realistic learning-theoretic approach. Hence, our scheme can be deployed for real-time operation without assuming acausal information on channel realizations, data/energy arrivals as required by offline optimization, nor does it rely on precise statistics of the system processes as is the case with online optimization. We formulate the problem as a partially observable identical payoff stochastic game (PO-IPSG) with factored controllers, in which the power control policy of each relay is adaptive to its channel and energy states as well as to the state of the source buffer. We equip each relay with a reinforcement learning procedure, and prove that the parallel execution of this procedure is convergent to (at least) a locally optimal solution of the formulated PO-IPSG. The proposed algorithm operates without explicit message exchange between the relays, while inducing only little source-relay signaling overhead. By simulation, we contrast the delay performance of the proposed method against existing heuristics for throughput maximization. It is shown that compared with these heuristics, the systematic approach adopted in this paper has a smaller sub-optimality gap once evaluated against a centralized optimal policy armed with perfect statistics.

*Index Terms*— bursty traffic, cooperative relaying, energy harvesting, power control, reinforcement learning, stochastic game, wireless communication.

## I. INTRODUCTION

COOPERATIVE relaying is a promising paradigm which results in broader coverage and in combating the wireless channel impairments. Relay-assisted transmission mitigates the need to use a high power at the transmitter, leading to prolonged battery life and lower level of interference [1]. Relays in wireless networks can be classified as decode-and-forward (DaF) relays, which decode and possibly re-encode the information before forwarding it, and amplify-and-forward (AaF) relays, which forward an amplified version of the signal without hard decoding. AaF relays compared with other types which require signal detection, are less complicated, have lower implementation cost, and are thus utilizable widely [4]. While cooperative relaying results in higher network capacity, in forwarding to the destination a representation of the signal it has received from the source, a relay consumes its own energy. Since replacing batteries for such devices is either impracticable or costly in several scenarios, recent advances in energy harvesting devices [5] have paved the way for self-sustainable relays [6] that power themselves from theoretically unlimited energy sources that are present in their surrounding environment (e.g., in the form of solar, vibration, thermoelectricity, etc.). However, the harvested energy rates are typically quite low with sporadic arrivals in random limited amounts, and it is thus desirable to accumulate the harvested energy by storing it in a buffer such as a rechargeable battery for subsequent usage. In practice, the energy buffer is restricted in size, and thus EH relays may face power outage whenever the energy consumption rate is higher than the harvesting rate. Hence, there is a need for novel power-use policies which exploit available information on the energy, channel and data arrival processes to efficiently utilize the harvested power for meeting application-specific demands.

### A. Literature Review

Exploiting both energy harvesting and cooperative communications has received a considerable interest recently [7-20]. The use of EH relays in cooperative communication was first introduced in [8], where a comprehensive performance analysis was conducted for relay selection and transmission power setting in an AaF network in terms of symbol error probability by using a probabilistic energy model. However, the results in [8] are mostly of analytical

V. Hakami is with the Department of Computer Engineering, Iran University of Science and Technology (IUST), Tehran 16846-13114, Iran (email: vhakami@iust.ac.ir).
M. Dehghan is with the Department of Computer Engineering and Information Technology, Amirkabir University of Technology (AUT), Tehran 15916-34311, Iran (e-mail: dehghan@aut.ac.ir).

interest rather than proposing a practical optimization scheme. More recently, several studies have come up with transmission control strategies (e.g., power allocation, relay selection, etc.) to optimize different network utility functions in EH relay systems [7,9,10,11,13,14,15,17,18,19,20,35]. These schemes can be categorized based on two main distinguishing features:

- *Optimization method (offline/online/learning-theoretic):* In offline optimization, it is assumed that all the future realizations of data/energy arrivals as well as the channel variations are known acausally before the system starts. In general, offline optimization problems are modeled as a mathematical program and the solution obtained can be considered as an upper bound on the performance of the actually stochastic system. In contrast, online optimization is much more realistic in the sense that only statistical knowledge but causal information on the realizations of the system states is assumed. A systematic way to approach online optimization is to formulate the problem as a stochastic dynamic program (DP) [21], and optimize the expected value of the long-run system performance. Nonetheless, in many practical scenarios either the characteristics of the channel variations and energy/data arrival processes change over time, or it is not possible to have reliable statistical information about these processes before node deployments. For example, in a sensor field with solar EH nodes distributed over a forest, each node's solar EH profile will depend on its location, and is subject to change based on the time of the day or the day of the week. To adapt the transmission scheme in real time, one should resort to learning-theoretic schemes as they are capable of converging to optimal transmission policy over time in the absence of prior knowledge on the statistics of the processes governing the communication system.

- *Traffic type assumption (saturated/bursty):* Under saturated traffic assumption, there are infinite data backlogs at the source, and the optimization objective is to improve the physical layer performance (e.g., throughput, outage probability or symbol error rate), by only accounting for channel and energy state processes. When traffic is bursty, however, there is a need for a buffer where packets can be queued. The "emptying" rate of the buffer becomes then the "service" rate. A physical-layer model that only captures the variation of the channel and energy completely disregards this issue, and it can result in arbitrary long average waiting time of the packets at the source buffer. When the end-to-end delay is of interest, we need to track the source queue size that develop under bursty traffic generation, and the allocation of power at relays should control the service rate to achieve delay optimization at the source data link layer.

The majority of the studies on EH relay systems lie within the offline optimization framework, and assume non-bursty source traffic type [7,9,10,13,14,15,18,19,20]. In [10], the problem of optimal power control for throughput maximization in an SRD network (one source-destination pair and one relay) is formulated as a nonlinear program in an offline setting. Both source and relay are harvesting entities, and the relay operates in half-duplex mode using AaF protocol. A similar setup is considered in [7], but for the case that both source and relay nodes have their own data to transmit to the destination, and the optimization objective is to maximize the total throughput. Also, in [9], the transmit power is jointly optimized with relay selection to handle the case of multiple relays. In [13], source and relay power allocation is optimized for an SRD system with a full-duplex relay using DaF protocol. Half-duplex DaF relaying is considered in [14], where it is assumed that only the source node can harvest energy. The case where both source and relay are EH nodes is handled in [15,18], while [20] considers two parallel EH relays (the so called diamond relay channel [22]). It is also worth noting that technically, the multi-relay case can be deemed equivalent to the OFDM relay with individual power constraint in each subcarrier. Accordingly, the studies in [38] and [39] have proposed optimization schemes for data and energy cooperation in relay-enhanced OFDM systems.

Some studies [9,10,19] propose online throughput maximization for the case of saturated source traffic. In [19], for instance, a stochastic DP formulation is given for optimal online power allocation in the case of DaF relaying. In [10], the online power allocation problem is formulated as a Markov decision process (MDP) [23] and a computationally simple scheme is provided for the special case where power control at the nodes is limited to on-off switching. Again, within the context of saturated source traffic type, there has also been a recent study which utilizes a solar-data-driven stochastic energy harvesting model in an MDP-based design, and obtains the optimal DaF relay power control policy to minimize the long-term average symbol error rate [35]. Under a bursty on-off Markovian traffic assumption, the study in [11] addresses online relay scheduling for EH wireless sensor networks. The problem is formulated as a partially observable MDP (POMDP) [24] in which the source node has to choose between direct or cooperative transmission modes depending on its own available energy, the states of its energy harvesting and event generation processes, and using only partial knowledge of the relay's state.

Finally, in [17], a multi-source, single relay cooperative network is considered where the traffic at the source nodes is assumed to be bursty and the forwarding protocol used by the relay is DaF. The transmit power of all nodes is assumed to be contributed by both the conventional AC utility power and the renewable energy. A distributed learning algorithm is proposed to minimize the sum of the average delay of the data flows by dynamic power, rate and link selection control.

*B. Motivation, Contributions and Outline*

Most prior art in optimizing the performance of EH relay systems belong to the realm of offline optimization, and primarily deal with the didactic single relay scenario [7,10,11,13,14,15,18,19]. Also, the existing online schemes require explicit knowledge of the statistics of the system processes [9,10,11,19] and do not address the case of bursty traffic in general where the optimization of the queueing delay is necessary. Unlike [17], in this paper, we consider an EH cooperative relay system consisting of multiple AaF relays which are powered solely by an energy harvesting storage with limited capacity. The source node, on the other hand, has a continuous power supply and maintains a data buffer for the bursty traffic flow towards the destination.

We aim at proposing a learning-theoretic scheme to control the relays' power consumption for optimizing the long-run average delay experienced by the source packets. Ideally, the learning mechanism should be able to dynamically control the transmit power at the relays in adaptation to the source buffer state information (SBSI) as well as the global channel state information (CSI) and energy state information (ESI) of the relays. This calls for a principled design based on a centralized stochastic DP formulation. However, such scheme is already doomed by the curse of dimensionality due to the huge space of global CSI, global ESI, as well as the exponential growth of the number of joint action combinations with the number of relays involved. Moreover to gain access to the global state of the system, a centralized controller would induce heavy signaling overhead. Hence, it is way more practical to empower the relays with decentralized autonomy to make their own decisions based on immediate local feedbacks and partial observability of the system state (i.e., local CSI (LCSI) and local ESI (LESI)). These decisions are not trivial since each relay faces the uncertainty of the system state (channel, buffer, energy) and of the other relays' actions and observations. To tackle these complications, we come up with a decentralized low overhead solution by making the following contributions:

- We rigorously formulate the delay-optimal multi-relay power control problem as a partially observable identical payoff stochastic game (PO-IPSG) [25] that considers the abovementioned properties of the EH relay system. PO-IPSG is a stochastic process that is collectively controlled by a group of independent agents who lack a central view of the global system state. Nevertheless, these agents have a shared objective; i.e., they are all interested in optimizing the utility of the team as a whole. The process is decentralized because none of the agents can control the whole process, and neither of the agents has a full view of the global state. This readily corresponds to our setting in that we also assume all relays in the network collectively aim at minimizing the average number of packets waiting in the source buffer. Also, by making each relay's power control policy adaptive to a partial view of the system consisting of SBSI, its LCSI, and LESI, the formulated PO-IPSG can systematically trade off long-term energy-efficiency and delay performance.

- Given our PO-IPSG formulation, we propose a *distributed learning-theoretic power control* (DLTPC) algorithm that can be used by the relays to learn their power control play strategies in the absence of statistical knowledge regarding the dynamics of channel, traffic, and energy processes. We construct DLTPC by building on and extending the classical results for gradient-based optimization of MDPs [27,28] and PO-IPSGs [25]. We show that our algorithm harmonizes the relays' policies so that their collective behavior is provably convergent to (at least) a locally optimal solution of PO-IPSG. As it turns out, DLTPC is a particularly lightweight algorithm, and its updates on the control policy induce only little source-relay signaling overhead with no explicit message exchange between the relays.

- By simulation, we show the sub-optimality gap between DLTPC and an MDP-based optimal policy that is armed with perfect statistics. It is evidenced that DLTPC has a smaller performance margin with the centralized controller compared to existing suboptimal throughput-maximizers for EH AaF multi-relay systems (e.g., [9]).

The rest of the paper is organized as follows: In Section II, we present the system model along with the general characteristics of the channel, traffic, and energy harvesting processes we assume in this paper. In Section III, we give our PO-IPSG-based formulation of the multi-relay delay optimization problem. In Section IV, the DLTPC algorithm is proposed for convergence to a locally optimal solution of the formulated PO-IPSG. Section V is dedicated to the comparative evaluation of the DLTPC algorithm. The paper ends with a concluding epilogue.

## II. SYSTEM MODEL

In this section, we describe the two-hop relay communication system, as well as the channel, traffic, and energy harvesting models. As a notational convention, the time index appears as a subscript, while a relay's index is always a superscript. Bold symbols are used for non-scalars (i.e., vectors or sets) at the social level, collecting quantities across all relays. A symbol associated with an individual relay (be it a scalar, a vector, or a set) is never in bold.

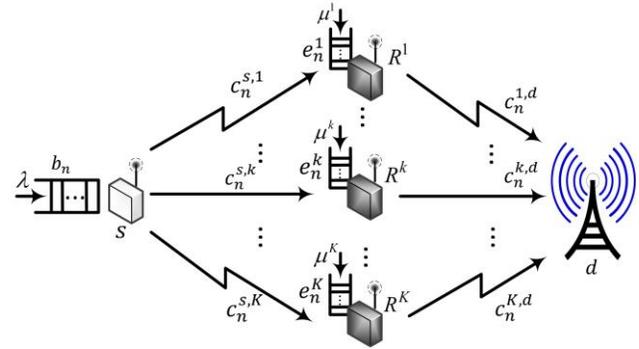

Fig. 1. A two-hop energy-harvesting cooperative relaying network.

### A. Energy-Harvesting Relay Communication System

The system under consideration is a two-hop relay network with one source node $s$, $K$ energy-harvesting relay terminals (each denoted by $R^k$, $k \in \mathcal{K} \triangleq \{1, \dots, K\}$) and one destination node $d$, as illustrated in Fig. 1. It is assumed that the source node's signal cannot reach the destination directly due to its limited transmission radius, and instead relies on the relays' assistance to transmit to $d$. We assume that all relays operate in half-duplex mode. A two-phase AaF protocol is used for $s$-to-$d$ packet delivery; more specifically, each time slot $n$ is split into two sub-slots, each with duration $\tau/2$. In the first sub-slot, the source broadcasts its own data with full transmission power $a^s$ to relay nodes. In the second sub-slot, according to the power control policy (defined in Section III.A and calculated by Algorithm 1), each relay decides whether to remain silent or to amplify the signal it has received from the source and forwards it to $d$. It is further assumed that the second hop transmissions by the relays are over orthogonal channels (e.g., using frequency division multiple access).

### B. Channel and Physical Layer Model

We consider a frequency non-selective block fading model, where $c^{s,k} \in \mathcal{C}^{s,k}$ denotes the channel fading gain from node $s$ to relay $R^k$. We use $\mathcal{C}^{s,k}$ to refer to the *local source-to-relay channel state information* (LSR-CSI) space; similarly, $c^{k,d} \in \mathcal{C}^{k,d}$ is used to denote the channel gain on the $R^k$-$d$ link, and $\mathcal{C}^{k,d}$ represents the *local relay-to-destination CSI* (LRD-CSI) space. We define the *local CSI* (LCSI) space for the $k$-th relay as $\mathcal{C}^k = \mathcal{C}^{s,k} \times \mathcal{C}^{k,d}$, where $c_n^k = \langle c_n^{s,k}, c_n^{k,d} \rangle \in \mathcal{C}^k$ is referred to as relay $R^k$'s LCSI at the $n$-th time slot. Also, we use $\mathcal{C} = \times_{k=1}^{K} \mathcal{C}^k$ to denote the space of the global CSI, collecting the channel gains across all the relays $R^k, k \in \mathcal{K}$.

**Assumption 1.** *The global CSI $\mathbf{c}_n = \langle c_n^k \rangle_{k \in \mathcal{K}} \in \mathcal{C}$ is quasi-static in each time slot. Furthermore, the process $\{\mathbf{c}_n\}_{n \in \mathbb{N}}$ is i.i.d. between slots with distribution $\mathbb{P}\{\mathbf{c}\}$. It is assumed that $\mathbb{P}\{\mathbf{c}\}$ is unknown and that each relay $R^k$ is only aware of its local CSI $c_n^k$ at time $n$, which can be estimated using channel reciprocity, assuming a time-division duplexing (TDD) system.* ■

Let $x$ represent the broadcast information symbol with unit energy from node $s$. The signal received by $R^k$ is given by:

$$y_n^{s,k} = \sqrt{a^s c_n^{s,k}}\, x + \eta,$$

where $\eta$ is the additive white Gaussian noise (AWGN). Without loss of generality, we assume that the noise power is the same over all links, denoted by $\sigma^2$. In phase 2, relay $R^k$ amplifies $y_n^{s,k}$, and forwards it to node $d$ with the chosen power $a_n^k \in \mathcal{A}^k$. The received signal $y_n^{k,d}$ at $d$ is as follows:

$$y_n^{k,d} = \sqrt{a_n^k c_n^{k,d}}\, x_n^{k,d} + \eta,$$

where, $x^{k,d}$ is the signal sent from $R^k$ to $d$, normalized to have unit energy; i.e., $x_n^{k,d} = \frac{y_n^{s,k}}{|y_n^{s,k}|}$.

Given the power profile $\mathbf{a}_n = \langle a_n^k \rangle_{k \in \mathcal{K}}$, the end-to-end AaF cooperative service rate is as [34]:

$$r_n^{s,\mathcal{K},d} = \gamma_L W \log_2 \left(1 + \frac{\sum_{k \in \mathcal{K}} \Gamma_n^{s,\mathcal{K},d}}{\Upsilon}\right), \quad (1)$$

where, $W$ is the bandwidth for transmission, $\gamma_L$ denotes a bandwidth factor which is set to 1 for energy-constrained settings, $\Upsilon$ is a constant denoting the capacity gap, and:

$$\Gamma_n^{s,\mathcal{K},d} = \frac{a_n^k a^s c_n^{s,k} c_n^{k,d}}{\sigma^2 (a^s c_n^{s,k} + a_n^k c_n^{k,d} + \sigma^2)}, \quad (2)$$

is the relayed signal-to-noise ratio (SNR) for source node $s$, which is helped by relay node $R^k$.

### C. Traffic Model and Source Buffer Dynamics

We assume there is one buffer at the source for the storage of packets. Let $l$ be the size of each packet and $A_n$ be the random new packet arrival at the $n$-th slot.

**Assumption 2.** *The arrival process $\{A_n\}_{n \in \mathbb{N}}$ is i.i.d. with distribution $\mathbb{P}\{A\}$ and mean $\lambda = \mathbb{E}[A]$. Also, packet arrivals occur at the end of each time slot. It is further assumed that the specific form of $\mathbb{P}\{A\}$ is unknown a priori.* ■

We use $b_n \in \mathcal{B}$ to denote the *source buffer state information* (SBSI), which is the number of packets in the source buffer at the beginning of the $n$-th time slot. $N_B$ denotes the maximum buffer size. When the buffer is full ($b_n = N_B$), new arrivals will be dropped. Finally, the buffer dynamics follow Lindley's equation (3):

$$b_{n+1} = \min\left(\left(b_n - \frac{\tau r_n^{s,\mathcal{K},d}}{2l}\right)^+ + A_n, N_B\right), \quad (3)$$

where $(.)^+$ stands for $\max(.,0)$.

### D. Energy Harvesting and Relay Energy Storage Dynamics

The energy harvesting process at each relay is modeled as a packet arrival process (e.g., see [37]) such that each energy packet is an integer multiple of a fundamental energy unit (EU). The relay $R^k$ is capable of harvesting a random number $H_n^k$ of energy packets from the environment at each time slot. The relay stores its harvested energy in its battery or a super-capacitor [26] with a finite capacity denoted by $N_E^k$ (energy packets), and all the energy harvested when the battery is full is lost. Also, the leakage within the battery or super-capacitor and the inefficiency in storing harvested energy are assumed to be negligible. Let $e_n^k \in \mathcal{E}^k$ be the amount of renewable energy in relay $R^k$'s energy storage at the beginning of the $n$-th time slot. We refer to $e_n^k$ as *local energy state information* (LESI). Also, we use $\mathcal{E} = \times_{k=1}^{K} \mathcal{E}^k$ to denote the space of the global ESI, collecting all possible LESI combinations across all the relays. Similarly, $\mathbf{e}_n = \langle e_n^k \rangle_{k \in \mathcal{K}} \in \mathcal{E}$ is referred to as the system's global ESI at the $n$-th time slot.

**Assumption 3.** *The arrival process $\{H_n^k\}_{n \in \mathbb{N}}, \forall k \in \mathcal{K}$ is i.i.d. with respect to $n$, and has distribution $\mathbb{P}\{H^k\}$ and mean $\mu^k = \mathbb{E}[H^k]$. We assume that the new energy arrivals are observed after the control actions are performed at each slot. It is assumed that $\mathbb{P}\{H^k\}$ and $\mathbb{E}[H^k]$ are unknown and each relay $R^k$ is only aware of its LESI $e_n^k$ at each time slot.* ■

Let $a_n^k$ denote the chosen power level by relay $R^k$ at time $n$. The LESI dynamics for each relay $R^k$ is as follows:

$$e_{n+1}^k = \min\left(e_n^k - a_n^k \frac{\tau}{2} + H_n^k, N_E^k\right). \quad (4)$$

where $a_n^k$ must satisfy the following energy availability constraint:

$$a_n^k \frac{\tau}{2} \le e_n^k, \forall k \in \mathcal{K}. \quad (5)$$

Finally, it is implicitly assumed that $a_n^k = 0$ means that relay $R^k$ remains inactive in time $n$.

### III. Problem Formulation

In this section, we formulate a decentralized power control policy for the relays to cooperatively optimize the average delay incurred by the source packets. In our system model, the dynamics of the source buffer depends, in part, on the packet arrival intensity $\lambda$, but it also depends on the cooperative service rate $r^{s,\mathcal{K},d}$ it receives from the relays, which is affected by their channel states as well as their energy harvesting profile. Accordingly, we define the power control policy at each relay to be adaptive to SBSI, as well as its LCSI and LESI. In particular, adaptation to LCSI is needed to opportunistically exploit the channel dynamics and gain more value for the power invested. SBSI-adaptability is needed to make the policy delay-aware under the conditions of unsaturated traffic and finite-length buffer at the source. Finally, given that the relays rely on energy harvesting for their operation, their control policies are subject to instantaneous energy availability constraints. An LESI-adaptive policy avoids inadvertent consumption of the harvested energy, and increases the odds that on urgent occasions a larger number of relays are available for rendering their service (i.e., higher diversity order), and they have more feasible power options at their disposal.

Our formulation is founded on the assumption that the relays would be working towards a common goal, i.e., the optimization of the incurred delay by the source packets. Altogether, our setup comes down to the coupled interaction of a number of agents with identical interest in a Markovian environment based on partial knowledge of the system state information and without explicit awareness of the action choices of the other agents. A systematic way to formulate this problem is to cast the system as a *partially observable identical payoff stochastic game* (PO-IPSG) [25]. We denote the PO-IPSG as a quintuple $\mathcal{G} = \langle \mathcal{K}, \mathcal{S}, \mathcal{A}, T, r \rangle$. $\mathcal{S} = \mathcal{B} \times \mathcal{C} \times \mathcal{E}$ is the global system state space, where each $s_n \in \mathcal{S}$ denotes the global system state at the $n$-th time slot, i.e., $s_n = \langle b_n, c_n, e_n \rangle$ consists of the SBSI, global CSI, and global ESI; likewise, we use $\mathcal{S}^k = \mathcal{B} \times \mathcal{C}^k \times \mathcal{E}^k$ to represent the space of partially observed system states from the viewpoint of relay $R^k$, $k \in \mathcal{K}$. Similarly, $s_n^k = \langle b_n, c_n^k, e_n^k \rangle$ denotes the $k$-th relay's observed state at the $n$-th time slot. $\mathcal{A}(e) = \times_{k=1}^{K} \mathcal{A}^k(e^k), \forall e \in \mathcal{E}$ is the battery state-dependent joint action space, i.e., different combinations of feasible power levels which can be chosen by the relays (see (5)). The mapping $T: \mathcal{S} \times \mathcal{A} \times \mathcal{S} \to [0,1]$ denotes the global state transition probabilities, and is discussed in more detail in III.B. Finally, $r: \mathcal{S} \times \mathcal{A} \times \mathcal{S} \to \mathbb{R}$ is the instantaneous reward function which is defined to be identical across all relays. More specifically, we define $r$ as a function of the number of vacant places in the source buffer; i.e.,

$$r(s_n, a_n, s_{n+1}) = \nu(N_B - b_{n+1}), \quad (6)$$

where $\nu$ is a positive constant. The dynamics of the game $\mathcal{G}$ proceeds as follows: at each time slot $n$, each relay $R^k$ observes its local state $s_n^k$ and selects an action $a_n^k$ according to its power control policy $u^k$ (to be specified in III.A). A composite action profile $a_n = \langle a_n^k \rangle_{k \in \mathcal{K}}$ from the joint action space $\mathcal{A}$ is executed, the system probabilistically transitions to the next state $s_{n+1}$ according to the law $T(s_{n+1}|s_n, a_n)$, and all relays receive the identical reward $r(s_n, a_n, s_{n+1})$. The system-wide objective is to maximize the *value of the game*, i.e., the long-run average of the received rewards.

### A. Factored Control Policy

We assume that the system is controlled by stationary policies. The stationarity of a policy implies that it depends on the history of the game only through the current state. Moreover, we parameterize the policy space by a set of continuous parameters $\Theta \in \mathbb{R}^\mathcal{D}$ of some dimension $\mathcal{D}$. In particular, as we are interested in decentralized optimization with partial state observability by the relays, we restrict ourselves to the space of *factored* joint controllers $\mathcal{U}^\Theta$, where each $u^\Theta \in \mathcal{U}^\Theta$ is a probabilistic mapping of the form $u^\Theta: \mathcal{S} \times \mathcal{A} \to [0,1]$ and it holds that $u^\Theta = \prod_{k=1}^{K} u^{\theta^k}$. Basically, $\Theta$ is defined to be the concatenation of individual relay policy parameters, i.e., $\Theta = \langle \theta^1, \dots, \theta^K \rangle$, and $u^{\theta^k}: \mathcal{S}^k \times \mathcal{A}^k \to [0,1]$ is relay $R^k$'s individual power control policy. $\theta^k$ is taken to be a $\mathcal{D}^k \triangleq |\mathcal{S}^k \times \mathcal{A}^k|$-dimensional vector of the form $\theta^k = \langle \theta_{s,a}^k \rangle_{s \in \mathcal{S}^k, a \in \mathcal{A}^k}$; i.e., the joint policy space is of dimension $\mathcal{D} = \sum_{k=1}^{K} \mathcal{D}^k$.

*Remark 1*: The factorization of action choice allows for parallel computation of the control policy by the relays as stated in Theorem 2 (Section IV). It also helps overcome the curse of dimensionality associated with the huge size of the joint state-action space $\mathcal{S} \times \mathcal{A}$; however, as argued in [25], a side-effect is that only a subset of policies from the full space of joint policies (corresponding to e.g., a central non-factored controller) can be represented. Hence, we can at best yield the best set of policies from within the restricted space $\mathcal{U}^\Theta$. ∎

A common way to express parametric policies in the literature (e.g., see [27]) is to assume a Gibbs-like distribution for the shape of $u^{\theta^k}(.)$; more precisely, the probability of choosing power level $a \in \mathcal{A}^k(e)$ by relay $R^k$ in state $s = \langle b, c, e \rangle \in \mathcal{S}^k$ is expressed as follows:

$$u^{\theta^k}(a|s) = \frac{\exp(\theta_{s,a})}{\sum_{\acute{a} \in \mathcal{A}^k(e)} \exp(\theta_{s,\acute{a}})}, \quad (7)$$

Note that the denominator in (7) is ensured to be non-zero by always having $a = 0$ as the feasible choice.

### B. State Transition Laws

Assume a joint parametric control policy $u^\Theta \in \mathcal{U}^\Theta$ is given. The probabilistic dynamics of the system state can be characterized in terms of $u^\Theta$ and the mapping T, which denotes the controlled transition probabilities; more specifically, we have:

$$\mathbb{P}\{s_{n+1}|s_n, u^\Theta(a_n|s_n)\} = T(s_{n+1}|s_n, a_n) u^\Theta(a_n|s_n), \quad (8)$$

where (recalling Assumption 1 on i.i.d. channels), we have:

$$T(s_{n+1}|s_n, a_n) = \mathbb{P}\{c_{n+1}\} \cdot T(b_{n+1}|s_n, a_n) T(e_{n+1}|e_n, a_n), \quad (9)$$

and the source buffer state transition is as follows:

$$T(b_{n+1}|s_n, a_n) = \begin{cases} \mathbb{P}\left\{A_n = b_{n+1} - \left(b_n - \frac{\tau r_n^{s,\mathcal{K},d}}{2l}\right)^+\right\}, & b_{n+1} < N_B \\ \sum_{A=N_B-\left(b_n - \frac{\tau r_n^{s,\mathcal{K},d}}{2l}\right)^+}^{\infty} \mathbb{P}\{A_n = A\}, & b_{n+1} = N_B \end{cases} \quad (10)$$

For the probabilistic transition of the global ESI, we have:

$$T(\boldsymbol{e}_{n+1}|\boldsymbol{e}_n, \boldsymbol{a}_n) = \prod_{k=1}^{\mathcal{K}} T^k(e_{n+1}^k|e_n^k, a_n^k),$$

where,

$$T^k(e_{n+1}^k|e_n^k, a_n^k) = \begin{cases} \mathbb{P}\left\{E_n^k = e_k^{n+1} - \left(e_n^k - \frac{\tau a_n^k}{2}\right)\right\}, & e_{n+1}^k < N_E^k \\ \sum_{E=N_E^k - \left(e_n^k - \frac{\tau a_n^k}{2}\right)}^{\infty} \mathbb{P}\{E_n^k = E\}, & e_{n+1}^k = N_E^k \end{cases} \quad (11)$$

### C. System-Wide Objective

As is common in infinite-horizon stochastic DP problems [21], we may seek policies that choose actions to optimize either the expected total discounted reward or the expected average-reward per step criterion. In this work, we opt for the time-averaged metric due to the following reasons:

- The average reward criterion puts more emphasis on the long-run performance of the system and does not discount its future behavior; without prior knowledge, each byte of a file or voice packet is of equal significance and it is hardly justified to discount later packets as inherently less important.
- Moreover, even if a formulation based on discounted-reward maximization is employed to trade off the delay experienced by recent and later packets, the discount factor needs to be chosen heuristically, which affects the performance of the derived power control policy.
- Finally, we set the goal in PO-IPSG $\mathcal{G}$ to be the maximization of the long-run average number of empty slots in the source buffer. As we clarify in the sequel (see Remark 3), this time-averaged metric in our problem is naturally related to the mean waiting time in the source buffer, and correlates well with an objective judgment of the system performance.

Now that we have stated our rationale for choosing a time-averaged criterion, in Remark 2, we impose a mild assumption on the set of admissible policies in order to ensure that the time-average criterion is well-defined:

*Remark 2:* Similar to other literature in MDP [12][28], we restrict our consideration to unichain policies in this paper. The stationary policy $\boldsymbol{u}^{\boldsymbol{\Theta}}$ is said to be unichain if the controlled Markov chain $\{s_n\}_{n\in\mathbb{N}}$ under $\boldsymbol{u}^{\boldsymbol{\Theta}}$ is ergodic [33]. In this case, $\{s_n\}_{n\in\mathbb{N}}$ has a unique steady state probability distribution $\boldsymbol{\pi}$, where for all $s \in \mathcal{S}$, $\pi(s) = \lim_{n\to\infty} \mathbb{P}(s_n = s)$ [28]. Now, we may define the optimization objective as (12):

$$\max_{\boldsymbol{\Theta}} \bar{\mathcal{R}}(\boldsymbol{u}^{\boldsymbol{\Theta}}) \triangleq \lim_{N\to\infty} \frac{1}{N}\sum_{n=0}^{N-1} \mathbb{E}^{\boldsymbol{u}^{\boldsymbol{\Theta}}}\{r_n\} = \mathbb{E}^{\boldsymbol{\pi}}\{\nu(N_B - b)\}. \quad (12)$$

where the $\mathbb{E}^{\boldsymbol{\pi}}$ denotes expectation w.r.t. the underlying probability $\boldsymbol{\pi}$. ∎

*Remark 3:* We have from the *extended Little's law* (c.f., Lemma 1, [30]) that the long-run average delay $\bar{\mathcal{D}}(\boldsymbol{u}^{\boldsymbol{\Theta}})$ of the source packets under the (unichain) policy $\boldsymbol{u}^{\boldsymbol{\Theta}}$ verifies the following inequality:

$$\bar{\mathcal{D}}(\boldsymbol{u}^{\boldsymbol{\Theta}}) \leq \lim_{N\to\infty} \frac{1}{N}\sum_{n=0}^{N-1} \frac{\mathbb{E}^{\boldsymbol{u}^{\boldsymbol{\Theta}}}\{b_n\}}{(1-\mathbb{P}_{drop})\lambda},$$

where $\mathbb{E}^{\boldsymbol{u}^{\boldsymbol{\Theta}}}$ is the expectation under stationary policy $\boldsymbol{u}^{\boldsymbol{\Theta}}$ and $\mathbb{P}_{drop}$ is the packet drop rate due to source buffer overflow. Here, we argue that since in practice, we target reasonable (e.g., 0.1%) drop rates, it holds that $\mathbb{P}_{drop} \ll 1$, and therefore the following is a good approximation for the average delay:

$$\bar{\mathcal{D}}(\boldsymbol{u}^{\boldsymbol{\Theta}}) \approx \lim_{N\to\infty} \frac{1}{N}\sum_{n=0}^{N-1} \frac{\mathbb{E}^{\boldsymbol{u}^{\boldsymbol{\Theta}}}\{b_n\}}{\lambda}.$$

Furthermore, this approximation is asymptotically tight as the data buffer size increases. Therefore, for sufficiently large buffer size and low load regime, maximizing $\bar{\mathcal{R}}(\boldsymbol{u}^{\boldsymbol{\Theta}})$ is a valid alternative to minimizing the average delay. ∎

**Definition 1** (*Local Optimal of PO-IPSG $\mathcal{G}$*). *A profile of power control policies $\boldsymbol{u}^{\boldsymbol{\Theta}^*} = \langle u^{\theta_1^*}, \dots, u^{\theta_K^*}\rangle \in \boldsymbol{\mathcal{U}}^{\boldsymbol{\Theta}}$ is the local optimal of the game $\mathcal{G}$ if it satisfies the following condition:*

$$\nabla_{\boldsymbol{\Theta}}\bar{\mathcal{R}}(\boldsymbol{u}^{\boldsymbol{\Theta}^*}) = \vec{\boldsymbol{0}}. \quad \blacksquare$$

**Theorem 1**. *The gradient in Definition 1 can be computed as (13)*:

$$\nabla_{\boldsymbol{\Theta}}\bar{\mathcal{R}}(\boldsymbol{u}^{\boldsymbol{\Theta}}) = \lim_{N\to\infty} \frac{1}{N}\sum_{n=0}^{N-1} \frac{\nabla_{\boldsymbol{\Theta}}\mathbb{P}\{s_{n+1}|s_n, \boldsymbol{u}^{\boldsymbol{\Theta}}(a_n|s_n)\}}{\mathbb{P}\{s_{n+1}|s_n, \boldsymbol{u}^{\boldsymbol{\Theta}}(a_n|s_n)\}} Q(s_n, a_n), \quad (13)$$

*where the function $Q(.,.)$ is the so-called differential reward function defined as follows:*

$$Q(\boldsymbol{x}, \boldsymbol{y}) = \lim_{N\to\infty} \mathbb{E}^{\boldsymbol{u}^{\boldsymbol{\Theta}}}\left\{\sum_{n=0}^{N-1}(r_n - \bar{\mathcal{R}}(\boldsymbol{u}^{\boldsymbol{\Theta}}))\Big|s_0 = \boldsymbol{x}, a_0 = \boldsymbol{y}\right\}. \quad (14)$$

*Proof.* The proof follows immediately from the derivation in [28, Section 3.2]. ∎

Note that (13) can be written in a more convenient form by realizing that:

$$\frac{\nabla_{\boldsymbol{\Theta}}\mathbb{P}\{s_{n+1}|s_n, \boldsymbol{u}^{\boldsymbol{\Theta}}(a_n|s_n)\}}{\mathbb{P}\{s_{n+1}|s_n, \boldsymbol{u}^{\boldsymbol{\Theta}}(a_n|s_n)\}} \\ = \nabla_{\boldsymbol{\Theta}}\ln[\mathbb{P}\{s_{n+1}|s_n, \boldsymbol{u}^{\boldsymbol{\Theta}}(a_n|s_n)\}] \\ = \nabla_{\boldsymbol{\Theta}}\ln[\boldsymbol{u}^{\boldsymbol{\Theta}}(a_n|s_n)]. \quad (15)$$

It is worth noting that a function such as $\nabla_{\boldsymbol{\Theta}}\ln[\boldsymbol{u}^{\boldsymbol{\Theta}}(a_n|s_n)]$, which is the gradient of a log-likelihood, is also known as a *score function* in classical statistics [31]. Finally,

$$\nabla_{\boldsymbol{\Theta}}\bar{\mathcal{R}}(\boldsymbol{u}^{\boldsymbol{\Theta}}) = \lim_{N\to\infty} \frac{1}{N}\sum_{n=0}^{N-1} \nabla_{\boldsymbol{\Theta}}\ln[\boldsymbol{u}^{\boldsymbol{\Theta}}(a_n|s_n)] Q(s_n, a_n), \quad (16)$$

In what follows, we present a distributed learning-theoretic procedure to steer the relays' behavior towards a delay-optimal power control policy $\boldsymbol{u}^{\boldsymbol{\Theta}^*}$ in the sense of Definition 1.

### IV. A MULTI-AGENT REINFORCEMENT LEARNING SOLUTION

In our PO-IPSG formulation, it is desired that the relays make coordinated decisions despite their independence of one another and despite their lack of omniscience (i.e., each single relay is unaware of the other relays' local states, and the policies they are pursuing). In order to harmonize the relays' behavior, in this section, we present a *distributed learning-theoretic power control* (DLTPC) algorithm to be executed in parallel by each relay involved.

In fully observable IPSGs, *value function-based learning* methods (e.g., [32]) have been proposed for discounted reward problems, which are convergent to the optimal Nash equilibrium. As for our PO-IPSG problem, however, we resort to *policy search* methods which have been shown to be a reasonable alternative to value-based methods for partially observable environments [36]. In particular, we follow the lead of Peshkin et al. in [25], which introduce a general method for using gradient ascent in multi-agent policy spaces to guarantee convergence to local optima (i.e., gradient zero operating points) of the game. Through a sketchy analysis, it has been shown in [25] that: when the search space is restricted to factored social policies $\mathcal{U}^{\Theta}$, joint gradient ascent performed by a central controller (with access to observation histories of the whole system) is equivalent to parallel gradient ascent performed by individual agents (with access only to their own partial view of the system history). Key to the argument in [25] is to show that:

I) The parallel algorithm samples gradients $\nabla_{\Theta}\bar{\mathcal{R}}$ from the correct distribution, and
II) The update increments used in gradient ascent are the same in the parallel algorithm as in the joint one.

Moreover, to satisfy these two conditions, an underlying requirement is that the agents perform synchronized updates on the estimates of their own components of the global gradient vector. Although the study in [25] is conducted in the context of discounted reward PO-IPSGs, but as we show in this paper, their line of argument can be extended to average-reward settings as well. However, the discussion in [25] is more of an outline lacking most details on the machinery of gradient estimation. We thus turn to standard techniques for estimation of the gradient of the average-reward in MDP literature [27][28]. These algorithms typically exploit the regenerative structure of the system' underlying Markov process to obtain unbiased gradient estimates based on the observations made in between regeneration times (i.e., between visits to a certain recurrent state). Applied to our PO-IPSG formulation, corresponding to every global regenerative cycle, we may define a local cycle for each relay during which it collects local observations to form an estimate of its own component of the global gradient vector. We show that at the expense of a very low signaling overhead, it can be arranged for the relays to agree on the termination of global regenerative cycles, thus satisfying the underlying requirement of synchronized updates in [25]. We then rigorously apply the line of argument in [25] to show that conditions I and II will be satisfied by our derivation (see Theorem 2 in Section IV). Based on this result, in Section IV.B, we discuss the update rules to be executed iteratively by each relay, and present DLTPC's pseudo code.

### A. Decentralized Computation of the Performance Gradient

Assume that the relay communication system is controlled via some factored joint parametric control policy $\boldsymbol{u}^{\Theta} \in \mathcal{U}^{\Theta}$ (c.f., Section III.A). The global system history is realized as an infinite-length trajectory of the form:

$$\boldsymbol{h}_{\infty} = [\boldsymbol{s}_0, \boldsymbol{a}_0, r_0, \boldsymbol{s}_1, \ldots, \boldsymbol{s}_{n-1}, \boldsymbol{a}_{n-1}, r_{n-1}, \boldsymbol{s}_n, \ldots]$$
$$\in \mathcal{H}_{\infty} \triangleq (\mathcal{S} \times \mathcal{A} \times \mathbb{R})^{\infty}.$$

Now, fix some $e^* \in \mathcal{E}^k, \forall k$ and let $\boldsymbol{e}^* \in \mathcal{E}$ be the global ESI where $e_n^k = e^*, \forall k$; likewise, fix some $b^* \in \mathcal{B}$. Finally, let $\mathcal{S}^* \triangleq \{\langle b^*, \boldsymbol{c}, \boldsymbol{e}^*\rangle, \forall \boldsymbol{c} \in \mathcal{C}\}$. With $\{\boldsymbol{s}_n\}_{n \in \mathbb{N}}$ being ergodic, elements of $\mathcal{S}^*$ recur infinitely often within any realization of the global system history. Let $t_m$ be the time of the $m$-th visit to $\mathcal{S}^*$. We refer to the following portion of history:

$$\boldsymbol{h}_m^* = [\boldsymbol{s}_{t_m}, \boldsymbol{a}_{t_m}, r_{t_m}, \boldsymbol{s}_{t_m+1}, \ldots, \boldsymbol{s}_{t_{m+1}-1}, \boldsymbol{a}_{t_{m+1}-1}, r_{t_{m+1}-1}, \boldsymbol{s}_{t_{m+1}}]$$

as the $m$-th *global renewal cycle* ($m \geq 1$). Under Assumption 1 for CSI and by *regenerative property* (e.g., see [29]), these pieces of system trajectory are i.i.d. We denote by $\ell(\boldsymbol{h}_m^*)$ the length of $\boldsymbol{h}_m^*$ that is equal to $\Delta t_m = t_{m+1} - t_m$. It is also convenient to introduce local versions of a renewal cycle observed through the prism of each relay $R^k$. In fact, corresponding to the $m$-th global renewal cycle $\boldsymbol{h}_m^*$, the relay $R^k$'s *local renewal cycle* is realized as follows:

$$h_m^{*,k} = [s_{t_m}^k, a_{t_m}^k, r_{t_m}, s_{t_m+1}^k, \ldots, s_{t_{m+1}-1}^k, a_{t_{m+1}-1}^k, r_{t_{m+1}-1}, s_{t_{m+1}}^k],$$

where, by definition of $t_m$, it holds that for all $k \in \mathcal{K}$: $s_{t_m}^k, s_{t_{m+1}}^k \in \mathcal{S}_k^* \triangleq \{\langle b^*, c^k, e^*\rangle, \forall c^k \in \mathcal{C}^k\}$; i.e., $h_m^{*,k}$ is of the same length as $\boldsymbol{h}_m^*$. Now, more generally, define $\mathcal{H}^*$ to be the space of all global renewal cycles; accordingly, $\mathcal{H}^{*,k}$ is used to refer to the space of all local renewal cycles for relay $R^k$. For $\boldsymbol{h}^* \in \mathcal{H}^*$, it holds that:

$$\mathbb{P}(\boldsymbol{h}^*|\Theta) = \prod_{n=0}^{\ell(\boldsymbol{h}^*)-1} T(\boldsymbol{s}_{[n+1,\boldsymbol{h}^*]}|\boldsymbol{s}_{[n,\boldsymbol{h}^*]}, \boldsymbol{a}_{[n,\boldsymbol{h}^*]})\boldsymbol{u}^{\Theta}(\boldsymbol{a}_{[n,\boldsymbol{h}^*]}|\boldsymbol{s}_{[n,\boldsymbol{h}^*]}). \quad (17)$$

where the notation $x_{[n,\boldsymbol{h}^*]}$ is used to refer to the component of $x$ realized at time $0 \leq n \leq \ell(\boldsymbol{h}^*)$ within $\boldsymbol{h}^*$. Now, by *renewal-reward theorem* (e.g., see [29]), the performance gradient $\nabla_{\Theta}\bar{\mathcal{R}}(\boldsymbol{u}^{\Theta})$ defined in (16) can be calculated as follows:

$$\nabla_{\Theta}\bar{\mathcal{R}}(\boldsymbol{u}^{\Theta}) = \frac{\mathbb{E}^{\boldsymbol{u}^{\Theta}}\left\{\sum_{n=0}^{\ell(\boldsymbol{h}^*)-1} \nabla_{\Theta} \ln[\boldsymbol{u}^{\Theta}(\boldsymbol{a}_{[n,\boldsymbol{h}^*]}|\boldsymbol{s}_{[n,\boldsymbol{h}^*]})]Q(\boldsymbol{s}_{[n,\boldsymbol{h}^*]}, \boldsymbol{a}_{[n,\boldsymbol{h}^*]})\right\}}{\mathbb{E}^{\boldsymbol{u}^{\Theta}}\{\ell(\boldsymbol{h}^*)\}}, \quad (18)$$

i.e., the expected total quantity earned during one cycle, normalized by the expected cycle duration. Similarly, the differential reward for $0 \leq n < \ell(\boldsymbol{h}^*)$ can be written as (19):

$$Q(\boldsymbol{x}, \boldsymbol{y}) = \mathbb{E}^{\boldsymbol{u}^{\Theta}}\left\{\sum_{j=n}^{\ell(\boldsymbol{h}^*)-1} \left(r_{[j,\boldsymbol{h}^*]} - \bar{\mathcal{R}}(\boldsymbol{u}^{\Theta})\right) \Big| \boldsymbol{s}_{[n,\boldsymbol{h}^*]} = \boldsymbol{x}, \boldsymbol{a}_{[n,\boldsymbol{h}^*]} = \boldsymbol{y}\right\}. \quad (19)$$

Replacing $Q$ with its estimate $\hat{Q}(\boldsymbol{s}_{[n,\boldsymbol{h}^*]}, \boldsymbol{a}_{[n,\boldsymbol{h}^*]}) \triangleq$

$\sum_{j=n}^{\ell(h^*)-1}\left(r_{[j,h^*]}-\bar{\mathcal{R}}(u^\Theta)\right)$ in (18), we have:

$$\overrightarrow{\nabla_\Theta^{\bar{\mathcal{R}}}} \triangleq \mathbb{E}^{u^\Theta}[\ell(h^*)]\nabla_\Theta \bar{\mathcal{R}}(u^\Theta) = \sum_{h^*\in\mathcal{H}^*} \mathbb{P}(h^*|\Theta) \times \left\{\sum_{n=0}^{\ell(h^*)-1}\nabla_\Theta \ln[u^\Theta(a_{[n,h^*]}|s_{[n,h^*]})]\hat{Q}(s_{[n,h^*]}, a_{[n,h^*]})\right\}, \quad (20)$$

where given that $\mathbb{E}^{u^\Theta}[\ell(h^*)]$ is a positive number, $\mathbb{E}^{u^\Theta}[\ell(h^*)]\nabla_\Theta \bar{\mathcal{R}}(u^\Theta)$ can be viewed as the expected gradient direction, and the zeroes of $\overrightarrow{\nabla_\Theta^{\bar{\mathcal{R}}}}$ are the same as those of $\nabla_\Theta \bar{\mathcal{R}}(u^\Theta)$.

Theorem 2 in the sequel establishes that the calculation of the direction of the performance gradient $\overrightarrow{\nabla_\Theta^{\bar{\mathcal{R}}}}$ can be done in a decentralized manner across the relays; i.e., each relay can independently calculate its individual gradient direction $\overrightarrow{\nabla_{\theta^k}^{\bar{\mathcal{R}}}}$ based on local information contained within its local renewal cycles $h^{*,k} \in \mathcal{H}^{*,k}$, and yet the ensemble of individual gradient directions recover the whole vector $\overrightarrow{\nabla_\Theta^{\bar{\mathcal{R}}}}$.

**Theorem 2.** *Assume $u^\Theta \in \mathcal{U}^\Theta$. The gradient direction $\overrightarrow{\nabla_\Theta^{\bar{\mathcal{R}}}}$ can be expressed as the vector:*

$$\overrightarrow{\nabla_\Theta^{\bar{\mathcal{R}}}} = \left\langle \overrightarrow{\nabla_{\theta^1}^{\bar{\mathcal{R}}}}, \ldots, \overrightarrow{\nabla_{\theta^K}^{\bar{\mathcal{R}}}} \right\rangle,$$

*in which each component $\overrightarrow{\nabla_{\theta^k}^{\bar{\mathcal{R}}}}, k \in \mathcal{K}$ is calculated as:*

$$\overrightarrow{\nabla_{\theta^k}^{\bar{\mathcal{R}}}} \triangleq \mathbb{E}^{u^\Theta}[\ell(h^*)]\nabla_{\theta^k}\bar{\mathcal{R}}(u^\Theta) = \sum_{h^{*,k}\in\mathcal{H}^{*,k}} \mathbb{P}(h^{*,k}|\Theta) \times \left\{\sum_{n=0}^{\ell(h^{*,k})-1}\nabla_{\theta^k}\ln\left[u^{\theta^k}\left(a_{[n,h^{*,k}]}\big|s_{[n,h^{*,k}]}\right)\right]\hat{Q}\left(s_{[n,h^{*,k}]}, a_{[n,h^{*,k}]}\right)\right\}, \quad (21)$$

*and,*

$$\hat{Q}\left(s_{[n,h^{*,k}]}, a_{[n,h^{*,k}]}\right) \triangleq \sum_{j=n}^{\ell(h^{*,k})-1}\left(r_{[j,h^{*,k}]}-\bar{\mathcal{R}}(u^\Theta)\right). \quad (22)$$

*Proof.* Please see Appendix A. ∎

In essence, Theorem 2 states that: If at each renewal cycle, all relays $R^k, k \in \mathcal{K}$ update their policy parameters $\theta^k$ along the gradient direction sampled from their distribution $\mathbb{P}(h^{*,k}|\Theta)$ in parallel, the parameter vector $\Theta$ gets updated along the gradient direction sampled from $\mathbb{P}(h^* = \langle h^{*,1}, \ldots, h^{*,K}\rangle|\Theta)$; i.e., the distributed algorithm is sampling from the correct distribution. Also, due to factorization, the update increments $\overrightarrow{\nabla_{\theta^k}^{\bar{\mathcal{R}}}}$ to be used in relay $R^k$'s gradient ascent are independent of the parameters in other relays' policies. Hence, the policy learning and control can be distributed among relays without requiring that they be informed of each others' states and choices of actions.

### B. Distributed Learning-Theoretic Power Control (DLTPC)

In this section, we present DLTPC (Algorithm 1), our distributed learning-theoretic power control scheme, which can lead the relays' collective behavior to a locally optimal delay performance. DLTPC relies on sample estimates of the performance gradient obtained during the actual system run-time to perform gradient-ascent in policy space. Hence, our algorithm does not need the explicit knowledge of the CSI, SBSI, and ESI statistics, and is an instance of model-free learning. This is as opposed to doing exact gradient-ascent, which requires the explicit knowledge of the transition laws T to analytically compute the gradient direction. In DLTPC, each relay updates its policy parameter $\theta_m^k$ at the end of each renewal cycle, i.e., between visits to $S^*$ (see (27) in Algorithm 1). To understand (27), note that according to (21) and (22), we can use:

$$F_m^k \triangleq \sum_{n=t_m}^{t_{m+1}-1}\frac{\partial \ln\left[u^{\theta^k}(a_n^k|s_n^k)\right]}{\partial \theta^k}\bigg|_{\theta^k=\theta_m^k} \sum_{j=n}^{t_{m+1}-1}\left(r_j-\bar{\mathcal{R}}(u^\Theta)\right), \quad (23)$$

as the $m$-th cycle estimate of $\overrightarrow{\nabla_{\theta^k}^{\bar{\mathcal{R}}}}$, which is obtained by each relay $R^k$ from the sample renewal cycle $h_m^{*,k}$. Now, to allow for more efficient recursive implementation of the summation (23) in Algorithm 1, we rewrite $F_m^k$ as follows:

$$F_m^k = \sum_{n=t_m}^{t_{m+1}-1}\left(r_n-\bar{\mathcal{R}}(u^\Theta)\right)\sum_{j=t_m}^{n}\frac{\partial \ln\left[u^{\theta^k}(a_n^k|s_n^k)\right]}{\partial \theta^k}\bigg|_{\theta^k=\theta_m^k}, \quad (24)$$

which makes it possible to incrementally construct $F_m^k$ using transient quantities $z_n^k$ and $g_n^k$ before reaching the end of each cycle. Accordingly, equation (27) in the pseudo-code is basically the standard rule for stochastic gradient–ascent in which the parameter $\alpha_m \in \mathbb{R}^+$ denotes a learning rate. Also, similarly to [27], $\bar{\mathcal{R}}(u^\Theta)$ in (24) is replaced via its estimate $\hat{\mathcal{R}}_m$, which is also updated at each renewal cycle via the recursion (25):

$$\hat{\mathcal{R}}_{m+1} := \hat{\mathcal{R}}_m + \alpha_m \sum_{n=t_m}^{t_{m+1}-1}\left(r_n-\hat{\mathcal{R}}_m\right). \quad (25)$$

Equation (25) is a stochastic approximation of the average reward $\bar{\mathcal{R}}(u^\Theta)$, and is consistent with the observation that for the $m$-th cycle, it holds:

$$\bar{\mathcal{R}}_{\Theta_m}(\approx \hat{\mathcal{R}}_m) = \frac{\mathbb{E}^{u^\Theta}\left\{\sum_{n=t_m}^{t_{m+1}-1}r_n\right\}}{\mathbb{E}^{u^\Theta}\{\Delta t_m\}}. \quad (26)$$

**Theorem 3.** *Choose $\alpha_m$ such that the sequence $\{\alpha_m\}$ be diminishing (i.e., $\alpha_m \xrightarrow{m\uparrow\infty} 0$), un-summable (i.e., $\sum_m \alpha_m = \infty$), but square summable (i.e., $\sum_m \alpha_m^2 < \infty$). Also, consider the sequence of parameters $\{\Theta_m\}$ generated by Algorithm 1. Then, $\{\hat{\mathcal{R}}_m\}$ converges (with probability 1), and the profile of power control policies $\{u^{\Theta_m}\}$ converges to the local optimal of PO-IPSG $\mathcal{G}$; i.e., $\nabla_\Theta \bar{\mathcal{R}}(u^{\Theta_m}) \xrightarrow{m\uparrow\infty} 0$ (w.p. 1).*

*Proof.* With this setup, DLTPC's update equations in (27)

and (28) are exactly along the lines of the single-agent iterates in ([27], Eqs. (15) and (16)); hence, the convergence of the gradient components (with respect to $\theta^k, \forall k$) of the performance measure $\bar{\mathcal{R}}(\boldsymbol{u}^{\boldsymbol{\Theta}_m})$ to zero can be established via the same arguments made in ([27], Proposition 3). Combine this with Theorem 2 to conclude. ∎

---

**Algorithm 1.** Distributed Learning-Theoretic Power Control

*Initialization*: Set iteration index $n := 0$, renewal cycle index $m := 0$, initial transient differential reward $\hat{Q}_0 := 0$, initial estimate for the average reward $\hat{\mathcal{R}}_0 := 0$; Initialize parameter vector $\theta_0^k$ randomly and set $z_0^k := \vec{0}, g_0^k := \vec{0}, \forall k \in \mathcal{K}$;

Source $s$ broadcasts data and its buffer state $b_0$;

**while** (TRUE)
  **for each** relay $k \in \mathcal{K}$ **do**
    1) Choose power $a_n^k \sim u_k^{\theta_m^k}(.\,|s_n^k)$;
    2) Transmit data to destination $d$ with *power* $a_n^k$;
    3) Inform $s$ only if battery level $e_{n+1}^k$ has reached $e^*$;
    4) Receive data from $s$ along with the next buffer state $b_{n+1}$, and the cycle termination signal $\sigma_n \overset{\text{def}}{=} \begin{cases} 1, & e_{n+1} = e^* \text{ and } b_{n+1} = b^* \\ 0, & default \end{cases}$;
    5) Update transient quantities for gradient and differential reward:
      // Calculate immediate reward:
      $r_n := \nu(N_B - b_{n+1})$;
      // Update the transient differential reward estimate:
      $\hat{Q}_{n+1} := \hat{Q}_n + (r_n - \hat{\mathcal{R}}_m)$;
      // Update the transient gradient estimate:
      $z_{n+1}^k := z_n^k + \left.\frac{\partial \ln[u^{\theta^k}(a_n^k|s_n^k)]}{\partial \theta^k}\right|_{\theta^k = \theta_m^k}$;
      $g_{n+1}^k := g_n^k + (r_n - \hat{\mathcal{R}}_m)z_{n+1}^k$;
    6) **if** $(\sigma_n == 1)$ // The end of the $m$-th renewal cycle
      // Update policy parameter:
      $\theta_{m+1}^k := \theta_m^k + \alpha_m g_{n+1}^k$;     (27)
      // Update the average reward estimate:
      $\hat{\mathcal{R}}_{m+1} := \hat{\mathcal{R}}_m + \alpha_m \hat{Q}_{n+1}$;     (28)
      // Reset transient quantities:
      $g_{n+1}^k = \vec{0}, \hat{Q}_{n+1} := 0, z_{n+1}^k = \vec{0}$;
      // Update the cycle index:
      $m := m + 1$;
    **end if**
  **end for**
  $n := n + 1$;     // Update the time index.
**end while**

---

### C. Discussion and Directions for Future Research

In this section, we give a few remarks about the underlying assumptions in this paper, and discuss how relaxing these assumptions can serve as a basis for future research.

The first issue has to do with our assumption on altruistic participation of the relays in forwarding the source signal. In fact, a relay's willingness to cooperate is taken for granted and our game-theoretic formulation is only a means to perform decentralized coordination and control and not a means of cooperation stimulation. A potential future direction, thus, includes extensions to systems with self-interested relaying terminals, where acquiring service from the relays requires an incentive mechanism.

The second issue is regarding the extension of our system model to the case where the source node also uses a state-dependent law to control its transmit power for minimizing the delay at its queue. While ideally, the source power should be treated as yet another "degree of freedom", we argue, however, that such extension is non-trivial as an adaptive source would induce non-stationary dynamics on the power adjustment procedure performed by the relays. In fact, proposing a systematic mechanism for jointly controlling the source and relays' power is beyond the scope of this paper since we cannot naively consider the source node as another player in our PO-IPSG formulation. Therefore, in Section II, we have explicitly restricted our system model to the case where the source is transmitting with a constant power supply (e.g., maximum allowed power). That being said, there exists, however, some fair justifications in support of our simplifying assumption: the source node in our system model does not rely on harvested energy but is instead connected to a fixed power supply. Also, no direct communication link is assumed between the source and the destination node. As such, it is fairly reasonable that the source can tap into its energy supply to power its transmission with little concern for replenishment of its energy budget. When the source node is a non-harvesting entity, there are several works in the context of EH relay systems where the source power is assumed fixed [8].

Finally, we need to discuss the case of buffer-aided relaying where the relay nodes have data queues as well. Cooperative networks with buffer-aided relays have the advantage that their achievable diversity is not bottlenecked by transmission order (unlike the stream-like communication in the conventional case where at each time slot, signal transmission starts from source and is then relayed to the destination) [41]. However, these relays may also incur larger packet delays which can be quite diverse for different packets. Hence, from the application point of view, the lack of a data buffer at the relays in our work can be justified by arguing that it is to advocate a simple relay design while also minimizing packet delay which is desirable in certain applications. There are also some technical complications in the way of extending the proposed approach to the case of relays with buffers: Reasonably enough, in buffer-aided relaying, it is typically the case that at each slot, only one relay is selected for either transmission or reception. This necessitates an explicit link selection mechanism which does not fit well with the collaborative all-playing nature of our PO-IPSG formulation and its identical-payoff structure. The systematic way to account for buffer-aided relaying is again a formulation based on stochastic dynamic programming; however, in order to come up with a realistic scalable solution, we need to take on a different approach for problem decomposition. There are some studies along this line (e.g., see [17]) which address delay optimization in the context of buffer-aided relaying by exploiting the structural properties inherent to the problem.

The setup considered in [17], however, only consists of a single relay which gives the problem a nice weakly coupled structure amenable to decomposition into sub-problems.

## V. PERFORMANCE EVALUATION

In this section, we evaluate the performance of our proposed DLTPC algorithm for decentralized power control in EH multi-relay systems. We compare DLTPC's performance with three other power control schemes:

(*a*) *Centralized MDP with perfect statistics*: we assume that an MDP controller exists which is aware of the probability distributions of the channel fading $\mathbb{P}\{c\}$, traffic arrival $\mathbb{P}\{A\}$, as well as the energy arrival processes $\mathbb{P}\{H^k\}$ for all relays $k \in \mathcal{K}$. Armed with this knowledge, one can use standard solution methods (e.g., relative value iteration [23]) to solve for an optimal joint power control policy $u: \mathcal{S} \to \mathcal{A}$, which maximizes an average reward measure defined similarly as (12). While in principle, this method can obtain superior performance compared to DLTPC, it suffers from both curses of dimensionality and modeling, and therefore has no practical relevance. However, the reward measure obtained using this procedure can serve as an upper bound against which to compare the DLTPC's performance.

(*b*) *Harvesting rate (HR) assisted scheme* [9]: The online-HR scheme proposed in [9] is a centralized online (suboptimal) algorithm for joint relay selection and power allocation in multi-relay AaF EH cooperative communication systems. However, unlike DLTPC, online-HR assumes infinite backlog at the source (saturated traffic assumption), and aims at maximizing the throughput. In order to make online decisions, the approach in [9] uses the causal information of ESI and CSI, but also needs the statistics of the harvesting and channel processes. The setup in [9] considers the case where the source node is also an EH entity; therefore, in our simulation, we remove this restriction and assume a continuous power supply for the source to make it comparable with DLTPC. At each slot, using the knowledge of mean harvesting rate and average channel SNRs, online-HR first determines the transmit power of the relays via a closed-form formula, and then a simple (centralized) optimization is solved to determine the relay with the maximum throughput.

(*c*) *Naive scheme* [9]: This algorithm is also centralized and online; however, it does not require the statistics of the harvesting and channel processes. At each time slot, the relays use their stored energies as their transmit powers. Using these transmit powers, the equivalent SNRs for all links are calculated. Then, the relay with the maximum equivalent SNR among all is selected to forward the signal to destination.

In what follows, we first compare the computational complexity of DLTPC with Online-HR and Online-Naive, and then present our numerical results in Section V.B.

### A. Comparison of Computational Complexity

At each time step, the Online-HR algorithm [9] has to compute the maximum system throughput achievable by every relay and then select the relay with the best value. Hence, its complexity is $(K)$ in each time step (i.e., linear in the number of relays). The Online-Naive algorithm has also the complexity of $O(K)$ per time step as it needs to select the relay which provides the maximum equivalent SNR among all the relays. Both these algorithms are centralized and need to gather global information from the whole network for their operations. On the other hand, our DLTPC is a particularly lightweight algorithm, working with minimal message signaling overhead between source and relays (see steps 3 and 4 in Algorithm 1). The algorithm's update rules are written in terms of efficient recursive formulae, which lead to negligible complexity. Also, if the policy function for each relay is chosen to have the convenient form in (7), the score function at step 5 can simply be calculated as:

$$\left.\frac{\partial \ln\left[u^{\theta^k}(a_n^k|s_n^k)\right]}{\partial \theta^k}\right|_{\theta^k=\theta_m^k} = \begin{cases} 1 - u^{\theta^k}(a|s)\Big|_{\theta^k=\theta_m^k}, & a = a_n^k, s = s_n^k \\ -u^{\theta^k}(a|s)\Big|_{\theta^k=\theta_m^k}, & a \neq a_n^k, s = s_n^k \\ 0, & s \neq s_n^k \end{cases}$$

Therefore, at each time step, DLTPC needs just a few standard algebraic operations, along with one random number generation to calculate the next action.

### B. Numerical Evaluation

We consider a setup with a total of $K = 8$ relays. The time slot duration is $\tau = 2$ms. We assume Poisson packet arrival with mean rate $\lambda$ pkt/ms, and the packet size is 1024 bytes. The total bandwidth is $W = 2.5$MHz. The source buffer is quantized to have 10 states (i.e., $N_B = 9$ pkts). Moreover, we assume that all relays harvest energy according to a Poisson energy arrival with mean rate $\mu^k = 0.25$ energy pkt/ms, $\forall k$, and the renewable energy is stored in a battery with maximum capacity $N_E^k = 4$ (energy pkts). The source transmission power is fixed at 5 (energy pkt/ms). Although our algorithm does not use the knowledge of the channel model, for the purpose of experiments, we simulate Rayleigh fading for each link. In this model, the channel states $c^{s,k}$ and $c^{k,d}$ ($\forall k$) are exponentially distributed random variables. However, as we consider a finite number of possible states, digital quantization is used to discretize the channel states. In particular, all the channel states are quantized into six probability bins with the boundaries specified as: {(-∞,-5.41 dB), [-5.41 dB,-1.59 dB), [-1.59 dB,-0.08 dB), [-0.08 dB,1.42 dB), [1.42 dB,3.18 dB), [3.18 dB,∞)}. Over these bins, the stochastic evolution of channel states is i.i.d. across time and independent across users. This discretization of channel states have been justified in [40]. We choose $\langle b^*, c, e^* \rangle = \langle N_B, ., (N_E^k)_k \rangle$ as the recurrent state marking the renewal cycles for DLTPC. Also, the initial learning rate is taken to be $\alpha_0 = 2.5 \times 10^{-4}$, and is diminished every 100 renewal cycles by a factor of 0.9.

Fig. 2 plots the progression of the average source buffer length over time under DLTPC along with the two other suboptimal policies. The mean data arrival rate is fixed at 2.0 pkt/ms. As can be seen, both the online-HR and online-naive schemes converge much more quickly, but are outperformed by DLTPC in the limit. In Fig. 3, we plot the policy of all relays (for one particular state-action pair) as the joint policy is driven towards the local optimal of the PO-IPSG.

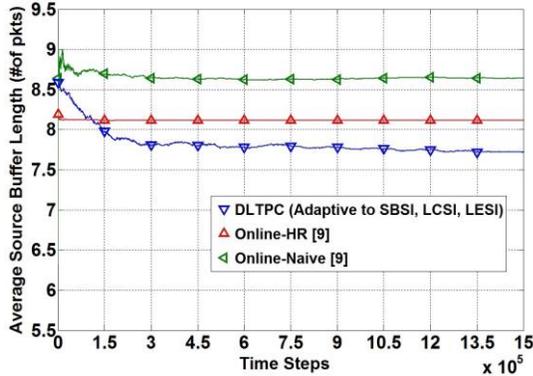
Fig. 2. Progression of average source buffer length.

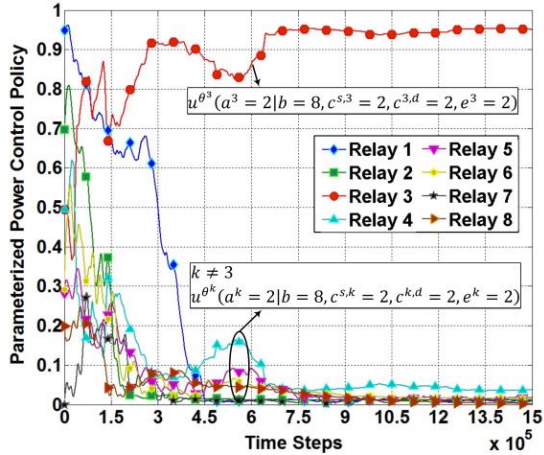
Fig. 3. Progression of power control policies.

Fig. 4 illustrates the average number of occupied slots in source buffer under various traffic intensities ($\lambda$ is varied from 1 pkt/ms to 2 pkt/ms). As a general trend, the source buffer gets more occupied as packet arrival rate increases. As expected, the MDP controller has the best performance gain among the four schemes. However, compared to the other two suboptimal policies, our SBSI-adaptive DLTPC algorithm maintains a smaller sub-optimality gap.

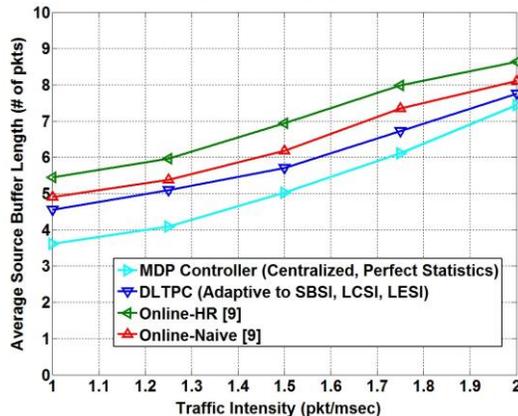
Fig. 4. The impact of input traffic intensity on delay performance.

Next, we investigate the impact of the relays' harvesting rate $\mu^k$ and battery capacity $N_E^k$ on delay performance. The mean Poisson data packet arrival rate is assumed to be 2.0 pkt/ms. In Fig. 5, we assume that the mean Poisson energy arrival rate for all relays is 0.25 energy pkt/ms, and plot the average number of occupied slots in source buffer for different values of battery size $N_E^k$ (from 4 to 8 energy pkts). The delay performance generally improves as battery capacity increases. However, DLTPC and online-HR can better exploit the enlarged energy storage with respect to the naive policy.

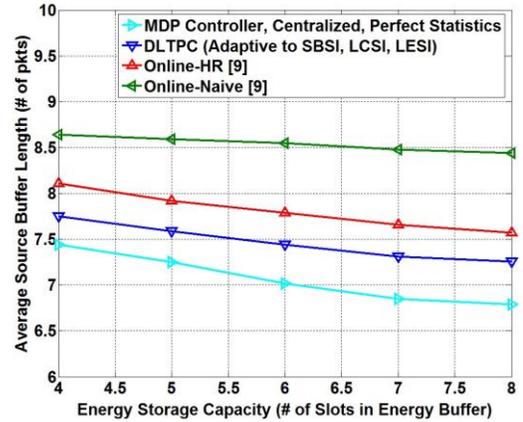
Fig. 5. The impact of energy storage capacity on delay performance.

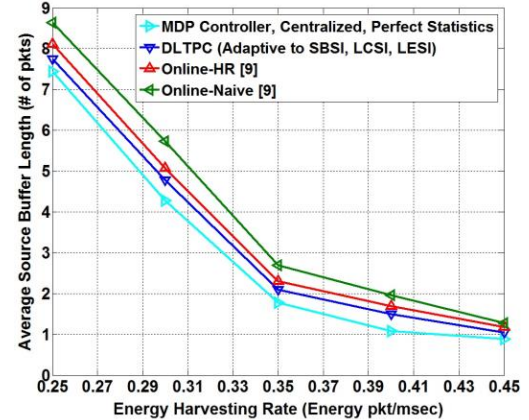
Fig. 6. The impact of energy harvesting rate on delay performance.

In Fig. 6, we fix the battery size $N_E^k$ to 4 energy packets, and instead vary the mean Poisson energy arrival rate for all relays from 0.25 to 0.45 energy pkt/ms. As expected, the source buffer receives a higher service rate as the relays' harvesting rate increases. In both plots, it is observed that our DLTPC algorithm maintains a better performance margin with respect to the centralized MDP controller.

## VI. CONCLUSION

The design of new protocols for cooperative networks with energy harvesting (EH) nodes is a promising research direction that incorporates cooperative benefits (diversity, capacity, etc.) with the energy harvesting concept. In pure EH relay systems, the nodes run on the energy harvested from the environment, and so are limited by their generation and storage capacities. This together with the stochastic nature of the profile of the harvested energy calls for the design of novel control policies which optimally utilize the power for meeting the application demands. However, the majority of the existing schemes have considered the case of single-relay SRD systems, and have focused on the optimization of the physical layer throughput by assuming non-bursty traffic arrival at the source. Also, the dominant methodologies for the optimization of these systems have been either offline optimizations assuming the availability of acausal information on the exact energy arrival instants and amounts, or online optimizations which rely on precise statistical knowledge of the system. In this paper, we considered an EH relaying system consisting of a bursty source with finite data buffer size whose transmission is cooperatively assisted by multiple EH relays. In order to optimize the average delay experienced by the source packets, we proposed a learning-theoretic solution which operates in the absence of prior knowledge of the statistics of the channel variation, traffic arrival and energy harvesting processes. The proposed method is highly decentralized and induces very low control overhead. Numerical evaluations demonstrated the superior delay performance of our solution compared to existing heuristics.

## APPENDIX A
### PROOF OF THEOREM 2

First, note that:

$$\hat{Q}(s_{[n,h^*]}, a_{[n,h^*]}) = \hat{Q}(s_{[n,h^{*,k}]}, a_{[n,h^{*,k}]}). \quad (29)$$

Now, by substituting $u^\Theta = \prod_{i=1}^K u^{\theta^i}$ in (20), it holds that:

$$\mathbb{E}^{u^\Theta}[\ell(h^*)]\nabla_{\theta^k}\bar{\mathcal{R}}(u^\Theta) = \sum_{h^* \in \mathcal{H}^*} \mathbb{P}(h^*|\Theta) \times$$

$$\left\{ \sum_{n=0}^{\ell(h^{*,k})-1} \nabla_{\theta^k} \ln\left[\prod_{i=1}^K u^{\theta^i}\left(a_{[n,h^{*,i}]}\big|s_{[n,h^{*,i}]}\right)\right] \hat{Q}\left(s_{[n,h^{*,k}]}, a_{[n,h^{*,k}]}\right) \right\} \quad (30)$$

$$= \sum_{h^* \in \mathcal{H}^*} \mathbb{P}(h^*|\Theta) \times$$

$$\left\{ \sum_{n=0}^{\ell(h^{*,k})-1} \left[\sum_{i=1}^K \nabla_{\theta^k} \ln\left[u^{\theta^i}\left(a_{[n,h^{*,i}]}\big|s_{[n,h^{*,i}]}\right)\right]\right] \hat{Q}\left(s_{[n,h^{*,k}]}, a_{[n,h^{*,k}]}\right) \right\} \quad (31)$$

$$= \sum_{h^* \in \mathcal{H}^*} \mathbb{P}(h^*|\Theta) \times$$

$$\left\{ \sum_{n=0}^{\ell(h^{*,k})-1} \nabla_{\theta^k} \ln\left[u^{\theta^k}\left(a_{[n,h^{*,k}]}\big|s_{[n,h^{*,k}]}\right)\right] \hat{Q}\left(s_{[n,h^{*,k}]}, a_{[n,h^{*,k}]}\right) \right\}, \quad (32)$$

Where the last equality is due to $\nabla_{\theta^k} \ln\left[u^{\theta^i}\left(a_{[n,h^{*,i}]}\big|s_{[n,h^{*,i}]}\right)\right] = 0$ for all $i \neq k$. Now, the entire term within the curly brackets in (32) can be written as a function $\phi(.)$ of relay $k$'s local renewal cycle $h^{*,k}$; i.e., $\phi(h^{*,k}) \triangleq$

$$\left\{ \sum_{n=0}^{\ell(h^{*,k})-1} \nabla_{\theta^k} \ln\left[u^{\theta^k}\left(a_{[n,h^{*,k}]}\big|s_{[n,h^{*,k}]}\right)\right] \hat{Q}\left(s_{[n,h^{*,k}]}, a_{[n,h^{*,k}]}\right) \right\}.$$

Also, given that the global renewal cycle $h^*$ can be described as the collection $\langle h^{*,1},\ldots,h^{*,K}\rangle$ of local renewal cycles across all relays, we have:

$$\sum_{h^* \in \mathcal{H}^*} \mathbb{P}(h^*|\Theta)\phi(h^{*,k})$$

$$= \sum_{\langle h_1^*,\ldots,h_K^*\rangle \in \mathcal{H}^*} \mathbb{P}(\langle h^{*,1},\ldots,h^{*,K}\rangle|\Theta)\phi(h^{*,k})$$

$$= \sum_{h_k^* \in \mathcal{H}_k^*} \left[\sum_{\langle h^{*,1},\ldots h^{*,k-1},h^{*,k+1},\ldots,h^{*,K}\rangle} \mathbb{P}(\langle h^{*,1},\ldots,h^{*,K}\rangle|\Theta)\right]\phi(h^{*,k})$$

$$= \sum_{h^{*,k} \in \mathcal{H}^{*,k}} \mathbb{P}(h^{*,k}|\Theta)\phi(h^{*,k}). \quad (33)$$

Hence, it follows that:

$$\overrightarrow{\nabla_{\theta^k}^{\bar{\mathcal{R}}}} = \sum_{h^{*,k} \in \mathcal{H}^{*,k}} \mathbb{P}(h^{*,k}|\Theta) \times$$

$$\left\{ \sum_{n=0}^{\ell(h^{*,k})-1} \nabla_{\theta^k} \ln\left[u^{\theta^k}\left(a_{[n,h^{*,k}]}\big|s_{[n,h^{*,k}]}\right)\right] \hat{Q}\left(s_{[n,h^{*,k}]}, a_{[n,h^{*,k}]}\right) \right\}. \quad (34)$$

∎

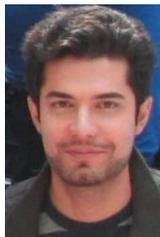

**V. Hakami** received his B.S. degree in computer engineering (software) and his M.S. and Ph.D. degrees in information technology (computer networking), all from Amirkabir University of Technology (AUT), Tehran, Iran, in 2004, 2008 and 2015, respectively. In 2016, He joined as an assistant professor to the Department of Computer Engineering, Iran University of Science and Technology (IUST), Tehran, Iran. His current research mainly focuses on cognitive control of computer networks using stochastic control theory, and game-theoretic learning.

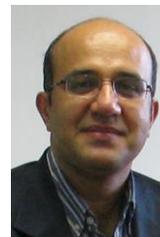

**M. Dehghan** (M'10) received his B.S. degree in computer engineering from Iran University of Science and Technology (IUST), Tehran, Iran, in 1992, and M.S. and Ph.D. degrees from Amirkabir University of Technology (AUT), Tehran, Iran, in 1995 and 2001, respectively. He is an associate professor of computer engineering and information technology at Amirkabir University of Technology (AUT). Before joining AUT in 2004, he was a research scientist at Iran Telecommunication Research Center (ITRC) working in the area of network quality-of-service and management. His research interests are in wireless networks, pattern recognition, fault-tolerant computing, and distributed systems.